\documentclass{aa}
\usepackage{graphics,epsf,epsfig,graphicx}

\begin{document}

\title{The large extent of dark matter haloes probed\\by the formation of tidal dwarf galaxies}

\author{Fr\'ed\'eric Bournaud \inst{1,2}, Pierre-Alain Duc \inst{1,3} \& Fr\'ed\'eric Masset \inst{1}}

\offprints{F. Bournaud, \email{Frederic.Bournaud@obspm.fr}}
\institute{
 CEA/DSM/DAPNIA, Service d'Astrophysique, Saclay, 91191 Gif sur Yvette Cedex, France
\and
   Ecole Normale Sup\'erieure, 45 rue d'Ulm, F-75005, Paris, France
\and
	CNRS FRE 2591}
\date{Accepted for publication in A\&A}
\authorrunning{F. Bournaud et al.}
\titlerunning{Dark matter haloes probed by the formation of tidal dwarf galaxies}

\abstract{In several interacting systems, gas accumulations as massive as 10$^9$ M$_{\sun}$ are observed near the tip of tidal tails, and are thought to be possible progenitors of Tidal 
Dwarf Galaxies. 
Using N-body simulations of galaxy interactions, we show that the existence of such features requires that dark matter haloes around spiral galaxies extend at least ten times further than the stellar disks. The massive gas clouds formed in our simulations have a kinematical origin and gravitationally collapse into dwarf galaxies that often survive for a few billion years.
}

\maketitle


\section{Introduction}

Dark matter haloes are known to surround galaxies, but whether they extend much further than visible matter is still debated. Standard cosmological models based on 
Cold Dark Matter theories predict that the dark halo of a galaxy should extend and maintain large circular velocities up to its virial radius, that is about 200 kpc for a galaxy of the mass of the
 Milky Way  (Navarro et al. 1996). This result is yet difficult to check by observations. Rotation 
curves of spiral galaxies can shed light on dark matter (DM) only at small radii and in the disk 
plane. 
At larger distances, on may however take advantage of the special location or morphology of 
some peculiar systems to probe the dark matter distribution. The kinematics of polar rings can give information on the flattening of dark haloes (Iodice et al. 2003).  The morphology of the long tidal 
tails (Springel \& White 1999; Dubinski et al. 1999) and the distribution and kinematics of 
satellite galaxies (Zaritsky \& White 
1994; Erickson  et al. 1999; Ibata et al.  2001) 
can be used to constrain the shape  of dark haloes. Finally one may study the 
internal dynamics of Tidal  Dwarf Galaxies (TDG)-- gravitationally bound objects formed out of 
tidal material (Duc et al., 2000) -- to  probe their  baryonic DM content and learn whether  part of 
the latter could originally have been located in a galactic disk (Braine et al. 2001).

In this letter, we show how the very existence of large accumulations of gaseous
tidal material, progenitors of massive TDGs,  can put some constraints on the large-scale extent 
of dark matter haloes. 
HI gas clouds with masses higher than $10^9$ M$_{\sun}$ are observed in the outer parts 
of the long tidal tails emanating from several nearby interacting systems
(e.g. Hibbard \& van Gorkom 1996; Duc et al. 1997,  2000; Nordgren et al. 1997; Braine et al. 2001)\footnote {See
 also in the HI Rogues Gallery (Hibbard et al. 2001) the collection of interacting doubles (available at  
http://www.nrao.edu/astrores/HIrogues/ )}. 
These clouds are usually real entities and not the results of projection effects. Indeed, they 
are able to convert their neutral hydrogen into molecular gas (Braine et al. 2001), form stars  
and may evolve into objects that have the apparent properties of dwarf galaxies. 

The formation of TDGs has been studied in numerical simulations showing the gravitational 
collapse of stellar clumps in tidal tails (Barnes \& Hernquist 1992), or the ejection of gas clouds
 from the parent galaxy (Elmegreen et al. 1993). These  objects have
 masses of 10$^7$ to 10$^8$ M$_{\sun}$  and are distributed all along the  tails like the 
TDG candidates found in numerous interacting systems (Weilbacher et al. 2000). However,
 the most 
massive TDG progenitors located near the tip of the tidal filaments have not yet been 
reproduced in  these simulations.

We present here new numerical simulations that aim at exploring two issues:
(1) how do HI clouds as massive as $10^9$ M$_{\sun}$  form in tidal tails ?
(2) are such clouds transient features that quickly fall back into their parent galaxy,
or are they the progenitors of long-lived objects able to contribute  significantly to the 
overall population of dwarf galaxies   and have a cosmological importance
(Hunter et al. 2000; Okazaki \& Taniguchi 2000) ? We will show that the answers to 
these questions also shed some light on the DM haloes of galaxies.


\section{Numerical simulations}
\subsection{Code and models}

We used N-body simulations of galactic encounters and mergers in order to  
study the response of gas to tidal interactions. The target galaxy consists of 50,000 particles 
  describing the stars, 100,000,  the dark matter, and 100,000 sticky ones, the gas. Its  stellar
 disk has  a 15 kpc radius, and is modeled by a Toomre disk with a scale-length of 7 kpc and a
 vertical scale-height of 1 kpc.
Its gaseous disk  has a radius of 2.3 times the stellar disk radius. The gas mass is a few 10$^9$ to 10$^{10}$ M$_{\sun}$ and the total luminous mass is 2  10$^{11}$ M$_{\sun}$. 
The second galaxy is gas free in our simulations, while in reality it may contain some gas. 
However, this does not affect our results  since the tidal tails of one galaxy are not disturbed by 
the gas present in the other one, except in very particular situations. 
 The dark matter is described as a standard isothermal sphere.
 For each galaxy, its distribution is computed to provide a flat rotation curve inside the halo 
truncation radius, which we chose between  3  and 10  times the stellar disk 
radius. 
 We also explored a range of mass ratios, relative velocities and impact
parameters  for the colliding galaxies. The characteristics of our different runs are listed
in Table 1.

The gravitational potential of stars and dark matter is computed via the three-dimensional FFT code presented in details in Bournaud \& Combes (2003). Using dark haloes as extended  as 150
  kpc  compels us to explore a large region, hence reducing the resolution of this 3-D code 
 to about 5 kpc. This resolution is large enough to compute the potential of stars and dark
 matter in which the gas response is studied, but is too low to reproduce the gas self-gravity inside 
the tails. 
We have therefore limited our study to coplanar galactic encounters, in which gas forms thin structures included in a plane. This enables the gas gravitational potential to be computed with a two-dimensional FFT code having a much higher resolution of about 150 pc (Bournaud \& Combes 2002). 
Once potentials are known, equations of motion are integrated by a leap-frog algorithm. The 
dissipative nature of the interstellar medium is described using the sticky-particles algorithm of 
Bournaud \& Combes (2002).

\subsection{Results}
In our first model presented in Figure~1, dark matter haloes do not extend much further than visible matter (stars and gas). Galaxy interaction
 makes a tidal tail develop, in which more than ten regions collapse under the effects of gravity. This 
result is very similar to what was obtained in previous simulations (e.g., Barnes \& Hernquist 1992): 
bound clumps of 10$^7$ to 10$^8$ M$_{\sun}$ are formed all along the tails. This model 
fails to reproduce the most massive gas accumulations seen in real systems. Even 
with varying the galaxies mass ratio, orbital parameters, and the initial extent and distribution of 
gas, such structures are never formed.

\begin{figure*}
   \includegraphics[width=10cm]{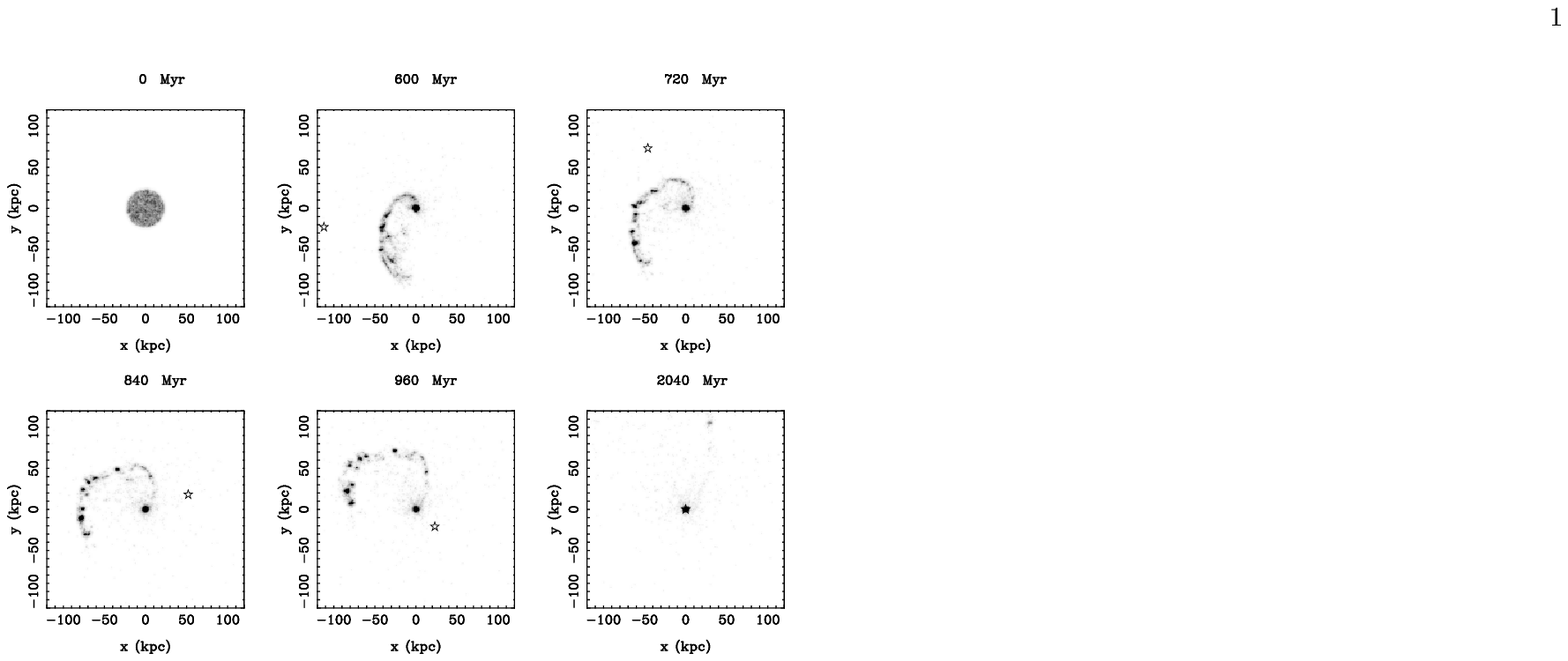}
   \caption{Response of the gaseous disk of a galaxy to a tidal interaction. The collision parameters are those of run 2 (see Table~1).
Dark haloes are initially truncated at three times the optical radius of the galaxies. The gas column-density is plotted in grayscale. The center of the second galaxy is represented by the star symbol.
More than ten gravitational clumps collapse along the tail with masses of 10$^7$ to 10$^8$ M$_{\sun}$.}
   \label{fig1}
\end{figure*}

 Figure~2 presents a  second model where the  dark haloes extend up to 10 times the radius of the stellar disk. 
The small tidal clumps  obtained with the first model are still there but, in addition, objects of 
10$^9$ M$_{\sun}$ may form in the outer parts of  tidal tails (see Fig.~3).
 They are  actually produced  in most of our runs (see Table~1), {\it provided that (1) long tidal 
tails are formed  (2)  the extent of haloes is at least a factor of ten greater than the radius of the stellar disks. } The long-lived tidal objects remain at more than 50 kpc from the galaxy center for at least
2 Gyr after their formation.

\begin{table}
\centering
\begin{tabular}{lcccl}
\hline
\hline
Run & $M$ & $V$ & $b$ & Results  for extended DMHs\\
\hline
 1 & 1  & 55 & 90 & 2 massive TDGs  \\
 2 & 2  & 55 & 90 & 1 massive TDG   \\
 3 & 3  & 55 & 90 &  no massive tidal object   \\
 4 & 0.5 & 55 & 90 & 1  massive TDG  \\
 5 & 0.25 & 55 & 90 &   1 massive TDG \\
 6 & 1  & 30 & 90 & 1 massive TDG   \\
 7 & 1  & 80 & 90 & 2 massive TDGs  \\
 8 & 1  & 125 & 90 &  no massive tidal object  \\
 9 & 1  & 55 & 65 &   2 massive TDGs \\
 10 & 1  & 55 & 120 & 1 massive TDG  \\
 11 & 1  & 55 & 150 &  1 massive TDG  \\
\hline
\end{tabular}
\caption{Parameters of the simulated collisions. $M$ is the mass ratio (disturbed 
galaxy to disturbing galaxy), $V$ the relative velocity at large distance (in km.s$^{-1}$), and 
$b$ the  impact parameter (in kpc).  The last column indicates whether, for simulations with a
large DM halo truncation radius,  massive TDGs were formed or not. In all these runs, orbital 
parameters are chosen to allow large tidal tails to develop. In particular the orbits of the 
galaxies were prograde.}
\label{runs}
\end{table}

\begin{figure*}
   \includegraphics[width=10cm]{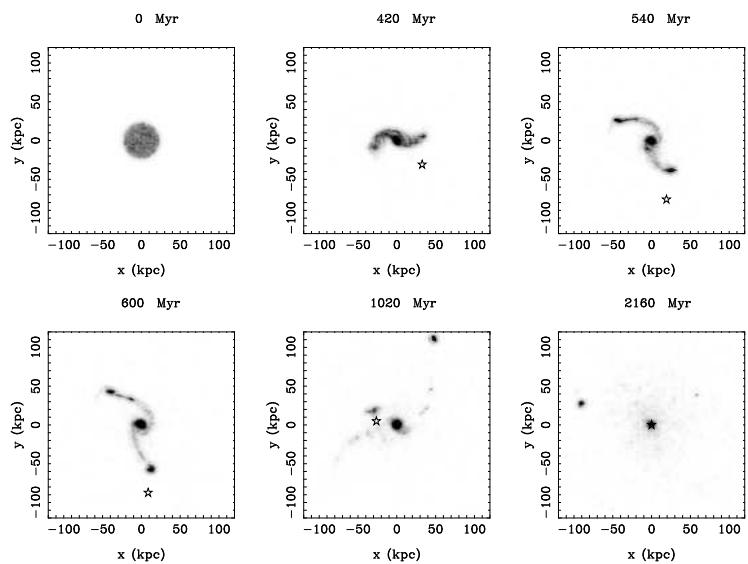}
   \caption{Same  run as in Fig.~1, with the truncation radius of the dark haloes extended up to 
10 times the radius of the stellar disks. Gas accumulations of 10$^9$ solar masses are formed at 
the  tip of each tidal tail. One of them quickly falls into the disturbing galaxy, while the other one forms a
 gravitationally bound object  that orbits at large radii on a nearly circular path.}
   \label{fig2}
\end{figure*}

\begin{figure}
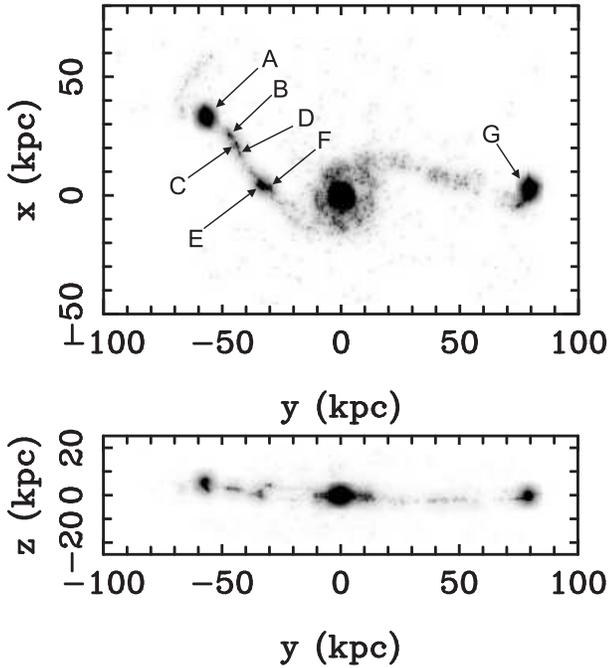

\centering
   \includegraphics[width=8cm]{Fg041_fig3a.eps}
   \includegraphics[width=8cm]{Fg041_fig3b.eps}
   \caption{Same run as in Fig.~2  observed at $t$=600 Myr. The initial disk plane is seen 
face-on (top) and edge-on (bottom). Two tidal dwarf galaxies (A and G) are formed near the 
extremity of the tails, with respective  masses of 1.1 and 1.3$\cdot$10$^9$ M$_{\sun}$. The 
smaller gravitational clumps distributed  along the tails (B to F) have masses from 3$\cdot$10$^7$ 
M$_{\sun}$ to 11$\cdot$10$^7$  M$_{\sun}$. The initial gas mass in the disk was 10$^{10}$ M$_{\sun}$.}
   \label{fig3}
\end{figure}

\section{Discussion and conclusions}\label{discu}

\subsection{Implications on dark haloes}
  Experiments probing the shape of  dark matter haloes  using the
kinematics  of satellite galaxies led  to somehow conflicting results (Zaritsky \& White, 1994;
Erickson et al., 1999) and to the suggestion that it was perhaps not universal. 
The massive and extended  dark matter haloes used in standard  CDM cosmologies 
have also been challenged by the  ability of colliding galaxies to form long tidal
tails (Mihos et al., 1998). In the N-body simulations by Springel \& White (1999), however, 
the more extensive haloes make stronger tails. 

 Our simulations indicate that the existence of very massive HI accumulations 
 in the outer parts of tidal tails, once formed, cannot be accounted for with 
isothermal dark haloes that are truncated at small radii, whatever the other physical parameters
 used in the simulations are. On the contrary, they are reproduced in most runs when the dark 
haloes of both interacting galaxies are  assumed to extend much further than the visible matter, 
at least 10 times or 150 kpc for a  galaxy of the mass of the Milky Way. 
The critical parameter for the  TDG-production ability is the halo size, not its mass. Indeed, 
we have run a set of simulations with haloes truncated at three times the stellar disk radius but as
massive as in the second model. They all failed at producing  massive tidal clouds. 


\subsection{Formation and survival of tidal dwarf galaxies} \label{evol}
 The formation of  bound star-forming clumps along tails is directly related to  gravitational 
instabilities in tidal structures  (Barnes \& Hernquist, 1992).  They are the progenitors 
of super star clusters and possibly to dwarf spheroidal galaxies  (Kroupa, 1998).

The more massive gas accumulations near the tip of the tidal tails -- those only obtained  
with extended dark  haloes -- seem to have a different origin. Indeed, they are still observed
in simulations where  the gas is not self-gravitating. 
 Thus, these structures are not initiated by self gravity but are firstly the result of a kinematical
 phenomenon. For parent galaxies
having both an extended halo, tidal forces make a significant fraction of the gas 
coming from beyond a critical distance in the parent's disk
to pile up near the extremity of the tail, whereas for a truncated halo, 
the gas is  spread all along the tail. 
Once enough material has accumulated, self-gravity leads to a local
collapse and ignites the star formation.
A more detailed investigation will be published in a forthcoming paper.

 The survival time of these TDGs will be determined by its ability to resist  tidal disruption,
internal starbursts, or an eventual merging with the parent galaxy. 
In our simulations, although much of the tidal material along the tails falls back 
 in less than  1 Gyr, the tidal dwarf galaxies that are formed at radii larger 
than 50 kpc are still observed after 2 Gyr. One third of them  remain on orbits that are almost 
circular (see Fig.~2) and  hence  appear as satellite galaxies.
Due to their low eccentricities, the final merging is postponed (Pe\~narrubia 
et al., 2002) and their life-time is increased. Therefore such tidal dwarf galaxies cannot 
be considered  as  transient objects. Old TDGs should exist (Hunter et al. 2000)
 and could actually represent a significant fraction of the dwarf galaxies 
present in nearby groups of galaxies, as suggested by 
Hunsberger et al.  (1996) or in the distant Universe, as 
speculated by Okazaki \& Taniguchi (2000).

We have restricted our study to coplanar encounters in order to treat the gas self-gravity with a 
large enough resolution. Since we found that the origin of the large tidal gas accumulations is 
not the
self-gravity, we have run a set of numerical experiments with non-coplanar encounters in which 
the gas is simply modeled by mass-less particles. The piling-up of gas  near the tail extremities 
still occurs only when dark haloes are extended 
enough.  The constraint on the size of dark haloes derived from the simulations appears 
therefore to be  robust. On the observational side, statistics of the number, location,
masses and ages of TDGs  need to be accumulated.


\begin{acknowledgements}
 We are grateful to Fran\c{c}oise Combes for her valuable comments on numerical techniques and results and  to Romain Teyssier for discussions on cosmological predictions for dark haloes.
 We thank the anonymous referee for his  careful reading of the manuscript and useful 
suggestions.
\end{acknowledgements}


\begin{thebibliography}{}


\bibitem[1]{1}Barnes, J. E., \& Hernquist, L. 1992, Nature, 360, 715
\bibitem[1]{2}Bournaud, F., \& Combes, F. 2002, A\&A, 392, 83
\bibitem[1]{3}Bournaud, F., \& Combes, F. 2003, A\&A, 401, 817
\bibitem[1]{4}Braine, J., Duc, P.-A., Lisenfeld, U., Charmandaris, V., Vallejo, O., et al. 2001, A\&A, 378, 51
\bibitem[1]{5}Dubinski, J., Mihos, J. C., \& Hernquist, L. 1999, ApJ, 526, 607
\bibitem[1]{6}Duc, P.-A., Brinks, E., Wink, J. E., \& Mirabel, I. F. 1997, A\&A, 323, 158
\bibitem[1]{7}Duc, P.-A., Brinks, E., Springel, V., Pichardo, B., Weilbacher, P., \& Mirabel, I. F. 2000, AJ, 120, 1238
\bibitem[1]{8}Elmegreen, B. G., Kaufman, M., \& Thomasson, M. 1993, ApJ, 412, 90
\bibitem[1]{9}Erickson, L., Gottesman, S. T., \& Hunter, J. H., Jr 1999, ApJ 515, 153
\bibitem[1]{10}Hibbard, J. E., \& van Gorkom, J. H. 1996, AJ, 111, 655
\bibitem[1]{99}Hibbard, J. E.,  van Gorkom, J. H., Rupen, M.P., \& Schiminovich, D. 2001, in
ASP Conf. Ser. 240, "Gas and Galaxy Evolution", J.E. Hibbard et al, eds 
\bibitem[1]{11}Hunsberger, S. D., Charlton, J. C., \& Zaritsky, D. 1996, ApJ, 462, 50
\bibitem[1]{12}Hunter, D. A., Hunsberger, S. D., \& Roye, E. W. 2000, ApJ, 542, 137
\bibitem[1]{13}Ibata, R, Lewis, G. F., Irwin, M., Totten, E., \& Quinn, T., 2001, ApJ,551, 294
\bibitem[1]{14}Iodice, E., Arnaboldi, M., Bournaud, F., Combes, F., Sparke, L. S., et al.  2003, ApJ, 585, 730
\bibitem[1]{15}Kroupa, P., 1998, MNRAS, 300, 200
\bibitem[1]{16}Mihos, J. C., Dubinski, J., \& Hernquist, L. 1998, ApJ, 494, 183 
\bibitem[1]{17}Navarro, J. F., Frenk, C. S., \& White, S. D. M. 1996, ApJ, 462, 563
\bibitem[1]{18}Nordgren, T. E., Chengalur, J. N., Salpeter, E. E., \& Terzian, Y. 1997, AJ, 114, 77
\bibitem[1]{19}Okazaki, T., \& Taniguchi, Y. 2000, ApJ, 543, 149
\bibitem[1]{20}Pe\~narrubia, J., Kroupa, P., \& Boily, C.M.  2002, MNRAS, 333, 779 
\bibitem[1]{21}Springel, V., \& White, S. D. M. 1999, MNRAS 307, 162
\bibitem[1]{22}Weilbacher, P. M., Duc, P.-A., Fritze v. Alvensleben, U., Martin, P., \& Fricke, K. J. 2000, A\&A, 358, 819
\bibitem[1]{23}Zaritsky, D., \& White, S. D. M. 1994, ApJ, 435, 599 

\end{thebibliography}
\end{document}